\documentstyle[manuscript,prd,aps]{revtex}
\newcommand{\gsim}{\mbox{\raisebox{-1.0ex}{$\stackrel{\textstyle >}
{\textstyle \sim}$ }}}
\newcommand{\lsim}{\mbox{\raisebox{-1.0ex}{$\stackrel{\textstyle <}
{\textstyle \sim}$ }}}
\newcommand{\bfx}{{\bf x}}

\newcommand{\bfk}{{\bf k}}
\newcommand{\bkp}{{\bf k'}}
\newcommand{\order}{{\cal O}}
\newcommand{\beq}{\begin{equation}}
\newcommand{\eeq}{\end{equation}}
\newcommand{\beqa}{\begin{eqnarray}}
\newcommand{\eeqa}{\end{eqnarray}}
\newcommand{\mpl}{M_{Pl}}
\newcommand{\lmk}{\left(}
\newcommand{\rmk}{\right)}
\newcommand{\lkk}{\left[}
\newcommand{\rkk}{\right]}
\newcommand{\lnk}{\left\{}

\newcommand{\call}{{\cal L}}
\newcommand{\calr}{{\cal R}}
\newcommand{\half}{\frac{1}{2}}
\newcommand{\kc}{\kappa\chi}
\newcommand{\bkc}{\beta\kappa\chi}
\newcommand{\gkc}{\gamma\kappa\chi}
\newcommand{\gbkc}{(\gamma-\beta)\kappa\chi}
\newcommand{\dchi}{\delta\chi}
\newcommand{\dphi}{\delta\phi}
\newcommand{\dOmega}{\delta\Omega}
\newcommand{\Phibd}{\Phi_{\rm BD}}
\newcommand{\echi}{\epsilon_\chi}
\newcommand{\ephi}{\epsilon_\phi}
\newcommand{\Phihat}{\hat{\Phi}}
\newcommand{\Psihat}{\hat{\Psi}}
\newcommand{\ahat}{\hat{a}}
\newcommand{\that}{\hat{t}}
\newcommand{\Hhat}{\hat{H}}
\newcommand{\zk}{z_k}
\begin{document}
\draft
\tighten
\title{Density fluctuations in Brans-Dicke inflation
\footnote{To be published in the proceedings for the
fourth workshop on general relativity and
gravitation edited by K.\ Maeda, T.\ Nakamura, and K.\ Nakao}}
\author{ Alexei A.~STAROBINSKY$^{1,2}$ and Jun'ichi YOKOYAMA$^{1}$}
\address{\hfill\\
$^1$Yukawa Institute for Theoretical Physics, Kyoto University, Uji
611 (Japan)\\
$^2$Landau Institute for Theoretical Physics, Kosygina St. 2, Moscow
117334 (Russia)\\}
\preprint{YITP/U-95-2}
\maketitle
\begin{abstract}
Spectrum of density perturbations in the Universe generated from
quantum-gravitational
fluctuations in slow-roll-over inflationary scenarios with the
Brans-Dicke gravity is calculated.  It is shown that after inflation
the isocurvature mode of perturbations may be neglected as compared
to the adiabatic mode, and that an amplitude of the latter
mode is not significantly different from that in the Einstein gravity.
However, the account of the isocurvature mode is necessary to obtain
the quantitatively correct spectrum of adiabatic perturbations.

\end{abstract}

\section{Introduction}

Inflationary expansion in the early Universe not only is able to
explain the observed degree of homogeneity, isotropy and flatness
of the present-day Universe \cite{inf}
but can also account for the origin of small initial density
perturbations which produce gravitationally bound objects
(galaxies, quasars, etc.) and the large-scale
structure of the Universe \cite{pert}.
Historically, among models of inflation
making use of a scalar field (called an inflaton), the original model or
the first-order phase transition model \cite{oriinf}
failed due to the graceful exit
problem, which was taken over by the new \cite{newinf}
and the chaotic \cite{chaoinf} inflation
scenarios where the inflaton scalar field is slow rolling during the
whole de Sitter (inflationary) stage. The latter property was shared
by the alternative scenario with higher-derivative quantum gravity
corrections
\cite{St80} (where the role of an inflaton is played by the Ricci
scalar $R$) just from the beginning. Note that a simplified version of
this scenario - the $R+R^2$ model - was even shown to be
mathematically
equivalent to some specific version of the chaotic scenario
\cite{W84} (see also a review in \cite{Got92}).
In order to obtain a small enough amplitude of density
perturbations in all these slow-roll-over models,
the inflaton should be extremely weakly coupled to other fields.
It is therefore not easy to find sound motivations to have a
such a scalar field in particle physics. (See, however,  \cite{MSYY}
for recent improvements along this line.)

Reflecting such a situation, the extended inflation \cite{extinf}
scenario was
proposed several years ago to revive a GUT Higgs field as the inflaton
by adopting non-Einstein gravity theories.  Although the first version
of the extended inflation model, which considers a first-order phase
transition in the Brans-Dicke theory \cite{BD},  resulted in failure
again due to the graceful-exit problem \cite{presc},
it triggered further study of more
generic class of inflation models in non-Einstein theories, in
particular extended chaotic inflation \cite{Lin90} where both
the inflaton and the Brans-Dicke scalar fields are in the slow
rolling regime during inflation. Note that the natural source of
Brans-Dicke-like theories of gravity is the low-energy limit of the
superstring theory \cite{FT85,C85} with the Brans-Dicke scalar
being the dilaton.

Several analyses have been done on the density
perturbations produced in extended new or chaotic inflation models
\cite{BM,Mc,MM,Der,Garcia}, all of which made use of
the constancy of the Bardeen's gauge-invariant quantity
$\zeta$ \cite{BST} or its equivalent on super-horizon scales
and match it directly to quantum field fluctuations at the moment
of horizon crossing which would be the correct procedure in a
single component inflationary model.
However, such a treatment is not correct in the
present case which contains two sources of quantum fluctuations,
namely, the inflaton and the Brans-Dicke scalar field.  In such a
model not only adiabatic but also isocurvature perturbations are
generated and $\zeta$ does not remain constant for all modes.

In the present article we analyze density perturbations generated in
both new and chaotic inflationary models in the Brans-Dicke gravity.
Keeping the above point in mind and extending the method used
in \cite{St85,PS} to find spectra of all modes of
adiabatic and isocurvature fluctuations in the
multiple inflationary scenario in the Einstein gravity (with scalar
fields interacting with each other through gravity only), we
separate adiabatic and isocurvature modes at the inflationary stage
and carefully evaluate their final amplitude.  As a result, we arrive at
a formula which is apparently very different from that used previously
but find that constraints on the model parameters in both models
remain approximately the same as in Einstein gravity if we express
them in terms of the values of coupling constants {\em at the end
of inflation}, contrary to the analysis by Berkin and Maeda \cite{BM}.

We make use of the conformal transformation which transforms the
original or the
Jordan frame to the Einstein frame in which equations are somewhat
simpler.  We calculate the spectrum of fluctuations in the Einstein
frame and then interpret the result to the Jordan frame.

The rest of the present article is organized as follows.  In \S 2 we
define the Lagrangian and perform conformal transformation to write
down the equations of motion in the Einstein frame.  Then in \S 3
solutions to these equations are given and they
are interpreted in the Jordan frame.  They are applied to both chaotic
and new inflationary models in \S 4.  Finally \S 5 is devoted
to discussion and conclusion.

\section{Basic equations}
\subsection{Conformal Transformation}

It has been shown that with the help of a conformal transformation a
wide class of non-Einstein gravity models can be recast in the action
\beq
S=\int\call\sqrt{-g}d^4x
=\int\lkk\frac{1}{2\kappa^2}\calr + \half g^{\mu\nu}\partial_\mu\chi
\partial_\nu\chi - U(\chi) + \half e^{-\gkc}g^{\mu\nu}
\partial_\mu\phi\partial_\nu\phi - e^{-\bkc}V(\phi)\rkk,
\label{action}
\eeq
where $\kappa^2=8\pi G$, $\beta$ and $\gamma$ are constants, $\chi$
and $\phi$ are the dilaton and the inflaton fields, respectively \cite{BM}.
For example, in the case of the Brans-Dicke theory \cite{BD}, the original
action in the Jordan frame
\beq
S=\int\lkk \frac{\Phibd}{16\pi}\hat{\calr} + \frac{\omega^2}{16\pi\Phibd}
\hat{g}^{\mu\nu}\partial_\mu\Phibd\partial_\nu\Phibd
+\half\hat{g}^{\mu\nu}\partial_\mu\phi\partial_\nu\phi
-V(\phi)\rkk\sqrt{-\hat{g}}d^4x,
\eeq
is transformed into (\ref{action}) through the conformal
transformation
\beq
g_{\mu\nu}=\Omega^2\hat{g}_{\mu\nu},~~~~~\Omega^2
\equiv\frac{\kappa^2}{8\pi}\Phibd \equiv
\exp\lmk\frac{\kc}{\sqrt{\omega+3/2}}\rmk,
\eeq
with
$\beta= 2\gamma=\frac{2}{\sqrt{\omega+3/2}}$
and $U(\chi)=0$.  Observations constrain $\omega$ to be $\omega >
500$ \cite{omega}, so that $\beta < 0.09 \ll 1$.  Since the Brans-Dicke field
$\Phibd$ remains practically constant in the radiation or matter
dominant stage (see below),
$\chi$ must be equal to zero at the end of inflation
in order to reproduce the correct value of the gravitational constant
today.  Although we focus on the original Brans-Dicke theory in the present
paper, we will not put the explicit values for $\beta$ and $\gamma$ below
until the very end of the calculation in order to sustain
applicability of our analysis to other non-Einstein gravity theories.

\subsection{Background equations in the Einstein frame}

Since the analysis is much simpler in the Einstein frame
(\ref{action}), we first investigate equations of motion derived from
(\ref{action}) and then transform the results into the original frame.
Taking the background  as the spatially flat
Friedmann-Robertson-Walker spacetime,
\beq
   ds^2=dt^2-a(t)^2d\bfx^2,
\eeq
the scalar field equations for the homogeneous parts read
\beqa
\ddot{\chi}+3H\dot{\chi}+\frac{\gamma\kappa}{2}e^{-\gkc}\dot{\phi}^2
-\beta\kappa e^{-\bkc}V(\phi)=0,  \label{chieqh}  \\
\ddot{\phi}+3H\dot{\phi}-\gamma\kappa\dot{\chi}\dot{\phi}+
e^{\gbkc}V'(\phi)=0,  \label{phieqh}
\eeqa
with an overdot denoting time derivation.
Because $\beta,~\gamma \ll 1$ and $V(\phi)$, being the inflaton's
potential, realizes slow roll over of $\phi$, the above equations can
be approximated as
\beqa
 3H\dot{\chi}&=&\beta\kappa e^{-\bkc}V(\phi),  \label{chislow} \\
 3H\dot{\phi}&=&-e^{\gbkc}V'(\phi),  \label{phislow}
\eeqa
with
\beq
  H^2=\lmk\frac{\dot{a}}{a}\rmk^2=\frac{\kappa^2}{3}e^{-\bkc}V(\phi),
  \label{hslow}
\eeq
during inflation.  We note that the above approximation is valid
provided that the following inequalities are satisfied.
\beqa
  \max\{ e^{-\gkc}\dot{\phi}^2,\dot{\chi}^2 \}&\ll&
  e^{-\bkc}V(\phi),\nonumber\\
  |V'(\phi)|&\ll& 3\kappa e^{-\gkc/2}V(\phi),  \label{slowrollcond} \\
  V''(\phi)e^{\gbkc}&\ll& H^2.  \nonumber
\eeqa
Then the system (\ref{chislow})--(\ref{hslow}) admits a solution as a
function of the scale factor as \cite{Lin90,BM}
\beq
  \chi  =\frac{\beta}{\kappa}\ln a \equiv -\frac{\beta}{\kappa}z,
  \label{chisol}
\eeq
\beq
  \kappa^2\int_{\phi_f}^{\phi}\frac{V(\varphi)}{V'(\varphi)}d\varphi
  =\frac{1-a^{\beta\gamma}}{\beta\gamma}
  =\frac{1-e^{-\beta\gamma z}}{\beta\gamma},  \label{phisol}
\eeq
where a subscript $f$ denotes the value of each quantity at the end of
inflation and we have set $a_f=1$.

In the case that $\phi$ changes slowly compared with $\chi$ and that
$V(\phi)$ can be regarded as a constant, evolution of the scale factor and
$\chi$ is given by
\beq
a(t)=a_0t^{\frac{2}{\beta^2}},~~~\chi(t)=\frac{1}{\kappa\beta}
\ln\lmk\frac{V(\phi)\kappa^2\beta^4t^2}{12}\rmk.
\eeq
We thus have a power-low inflation with an extremely large power exponent.

\subsection{Linear perturbations}

We now turn to linear perturbation, which is most easily taken into
account in a gauge-invariant manner using the longitudinal gauge,
\beq
  ds^2=(1+2\Phi)dt^2-a(t)^2(1-2\Psi)\delta_{ij}dx^i dx^j,
\label{metric}
\eeq
where $\Phi$ and $\Psi$ are gauge-invariant variables related with
Bardeen's $\Phi_A$ and $\Phi_H$ as $\Phi=\Phi_A$ and $\Psi=-\Phi_H$
\cite{Bardeen}.
Assuming an $e^{i\bfk\bfx}$ spatial dependence, each Fourier mode
satisfies the following equations of motion which are derived from the
perturbed Einstein equations.
\beq
  \Phi=\Psi,
\eeq
\beq
  \dot{\Phi}+H\Phi=\frac{\kappa^2}{2}(\dot{\chi}\dchi +
  e^{-\gkc}\dot{\phi}\dchi),
\eeq
\beqa
  \ddot{\dchi}+3H\dot{\dchi}
  +\lmk \frac{k^2}{a^2} -\frac{(\gamma\kappa)^2}{2}e^{-\gkc}\dot{\phi}^2
  +(\beta\kappa)^2e^{-\bkc}V(\phi)\rmk\dchi
  +\gamma\kappa e^{-\gkc}\dot{\phi}\dot{\dphi} \nonumber \\
  -\beta\kappa e^{-\bkc}V'(\phi)\dphi
  =2(\ddot{\chi}+3H\dot{\chi})\Phi+\dot{\Phi}\dot{\chi}
  +3\dot{\Psi}\dot{\chi}+\gamma\kappa e^{-\gkc}\dot{\phi}^2\Phi,
\eeqa
\beqa
  \ddot{\dphi}+(3H-\gamma\kappa\dot{\chi})\dot{\dphi}+
  \lmk  \frac{k^2}{a^2}+e^{\gbkc}V''(\phi)\rmk\dphi
  -\gamma\kappa\dot{\phi}\dot{\dchi}  \nonumber \\
  +(\gamma-\beta)\kappa V'(\phi)e^{(\gamma-\beta)\kappa\chi}\dchi
  =2(\ddot{\phi}+3H\dot{\phi})\Phi+\dot{\Phi}\dot{\phi}
  +3\dot{\Psi}\dot{\chi}-2\gamma\kappa\dot{\phi}\dot{\chi}\Phi,
\eeqa
where $\dchi$ and $\dphi$ are gauge invariant fluctuation variables of
the respective fields and we have suppressed the argument $\bfk$.
Under the slow-roll approximations (\ref{slowrollcond}) the last two
equations read
\beqa
  \ddot{\dchi}+3H\dot{\dchi}
  +\lmk \frac{k^2}{a^2}
  +(\beta\kappa)^2e^{-\bkc}V(\phi)\rmk\dchi
  +\gamma\kappa e^{-\gkc}\dot{\phi}\dot{\dphi}
  -\beta\kappa e^{-\bkc}V'(\phi)\dphi  \nonumber \\
  =2\beta\kappa e^{-\gkc}V(\phi)\Phi+\dot{\Phi}\dot{\chi}
  +3\dot{\Psi}\dot{\chi},  \label{chifluceq}
\eeqa
\beqa
  \ddot{\dphi}+3H\dot{\dphi}+
  \lmk  \frac{k^2}{a^2}+e^{\gbkc}V''(\phi)\rmk\dphi
  -\gamma\kappa\dot{\phi}\dot{\dchi}
  +(\gamma-\beta)\kappa V'(\phi)e^{(\gamma-\beta)\kappa\chi}\dchi \nonumber\\
  =-2e^{\gbkc}V'(\phi)\Phi+\dot{\Phi}\dot{\phi}
  +3\dot{\Psi}\dot{\chi}.  \label{phifluceq}
\eeqa

Since what we need are the non-decreasing adiabatic and isocurvature
modes on large scale $k \ll aH$, which turn out to be weakly
time-dependent as will be seen in the final result \cite{PS}, we may
consistently neglect $\dot{\Phi}$ and those terms containing two time
derivatives.  Then the equations of motion are greatly simplified to
\beq
  \Phi=\frac{\kappa^2}{2H}(\dot{\chi}\dchi + e^{-\gkc}\dot{\phi}\dphi)
  =\frac{\beta\kappa}{2}\dchi -\frac{V'(\phi)}{2V(\phi)}\dphi,  \label{lgr}
\eeq
\beq
  3H\dot{\dchi}+(\beta\kappa)^2e^{-\bkc}V(\phi)\dchi
  -\beta\kappa e^{-\bkc}V'(\phi)\dphi
  = 2\beta\kappa e^{-\bkc}V(\phi)\Phi,  \label{lchi}
\eeq
\beq
  3H\dot{\dphi}+e^{\gbkc}V''(\phi)\dphi
  + (\gamma-\beta)\kappa V'(\phi)e^{(\gamma-\beta)\kappa\chi}\dchi
  = -2e^{\gbkc}V'(\phi)\Phi.  \label{lphi}
\eeq

\section{Generation of perturbations}

\subsection{Solution of non-decreasing modes}

We shall now solve eqs.\ (\ref{lgr})--(\ref{lphi}).  Inserting
(\ref{lgr}) to (\ref{lchi}) we easily find $3H\dot{\dchi}=0$, or
\beq
  \dchi = \frac{\beta}{\kappa}Q_1 = {\rm const.}
\eeq
Similarly (\ref{lphi}) is transformed to
\beq
  3H\dot{\dphi} + e^{\gbkc}\lmk V''(\phi) -
\frac{V'(\phi)^2}{V(\phi)}\rmk \dphi + \beta\gamma
e^{\gbkc}V'(\phi)Q_1=0.  \label{lphi2}
\eeq
The above equation can be solved taking $\phi$ as an independent
variable instead of $t$.
With the help of (\ref{phislow}) we find
\beq
  3H\frac{d~}{dt}=3H\dot{\phi}\frac{d~}{d\phi}
  =-e^{\gbkc}V'(\phi)\frac{d~}{d\phi},
\eeq
so that (\ref{lphi2}) reads
\beq
  \frac{d~}{d\phi}\dphi=\lmk \frac{V''(\phi)}{V'(\phi)}
  -\frac{V'(\phi)}{V(\phi)}\rmk \dphi + \beta\gamma Q_1.
\eeq
Its solution is given by
\beq
  \dphi=\frac{V'(\phi)}{V(\phi)}
  \lmk \beta\gamma
  Q_1\int_{\phi_f}^{\phi}\frac{V(\varphi)}{V'(\varphi)}d\varphi
  + \frac{Q_2-Q_1}{\kappa^2} \rmk=-\frac{V'(\phi)}{\kappa^2V(\phi)}
  (Q_1 e^{\gkc}-Q_2),
\eeq
with $Q_2$ being another integration constant.
We therefore find
\beq
  \Phi=\frac{\beta^2}{2}Q_1
  +\frac{1}{2\kappa^2}\lmk\frac{V'(\phi)}{V(\phi)}\rmk^2
  (e^{\gkc}Q_1-Q_2).
\eeq
Thus we have obtained a generic solution of the system
(\ref{lgr})--(\ref{lphi}) containing two undetermined constants.

In order to clarify physical meaning of the above solution, we should
divide it to adiabatic and isocurvature modes.  The latter mode is
characterized by its vanishingly small contribution to the
gravitational potential $\Phi$, while the growing adiabatic mode can
be described  by the following universal expression.
\beqa
   \Phi&=&C_1\lmk 1-\frac{H}{a}\int_0^ta(t')dt'\rmk
   \cong -C_1\frac{\dot{H}}{H^2},  \label{gruniv}\\
   \frac{\dchi}{\dot{\chi}}&=&\frac{\dphi}{\dot{\phi}}=
   \frac{C_1}{a}\int_0^t a(t')dt'\cong \frac{C_1}{H}, \label{scalaruniv}
\eeqa
where $C_1$ is a constant and the latter approximate equality in each
expression is satisfied during the inflationary stage.

 From (\ref{chislow})--(\ref{hslow}) we find,
\beq
  \dot{\chi}=\frac{\beta}{\kappa}H,~~~
  \dot{\phi}=-H\frac{e^{\gkc}V'(\phi)}{\kappa^2V(\phi)},~~~
  \frac{\dot{H}}{H^2}=\frac{\beta^2}{2}+\frac{e^{\gkc}}{2\kappa^2}
  \lmk\frac{V'(\phi)}{V(\phi)}\rmk^2,
\eeq
and it turns out that defining new constants $C_1$ and $C_3$ by
$Q_1=C_1-C_3$ and $Q_2=-C_3$ discriminates between adiabatic and isocurvature
modes in the final result.  That is,
\beqa
\frac{\dchi}{\dot{\chi}}&=&\frac{C_1}{H}-\frac{C_3}{H},  \label{chifluc}\\
\frac{\dphi}{\dot{\phi}}&=&\frac{C_1}{H}+\frac{C_3}{H}(e^{-\gkc}-1),
\label{phifluc} \\
\Phi&=&-C_1\frac{\dot{H}}{H^2}
+C_3\lkk \frac{1}{2\kappa^2}(1-e^{\gkc})
 \lmk\frac{V'(\phi)}{V(\phi)}\rmk^2 - \frac{\beta^2}{2}\rkk.   \label{grfluc}
\eeqa
In the above expressions, terms in proportion to $C_1$ and $C_3$
represent adiabatic and isocurvature modes, respectively.  The
isocurvature nature of the $C_3-$terms is guaranteed by the fact that
the second term in the right-hand-side of (\ref{grfluc}) is
vanishingly small in the last stage of inflation when we have
$\chi \cong 0$.

\subsection{Quantum fluctuations}

We shall next determine the constants $C_1$ and $C_3$ from amplitudes
of quantum fluctuations of the scalar fields generated during the
inflationary stage.  Thanks to the inequalities (\ref{slowrollcond}),
eqs.\ (\ref{chifluceq}) and (\ref{phifluceq}) can be approximated by
equation of motion of a free massless scalar field in inflating
background for $k \geq aH$ and even in the region $k < aH$ but with
$H(t_k) \gg |\dot{H}(t_k)|(t-t_k)$ where $t_k$ is the time $k-$mode
leaves the Hubble horizon during inflation.
The standard quantization gives the well-known result, that is, the
Fourier components of the fields can be represented in the form
\beq
  \dchi (\bfk)=\frac{H(t_k)}{\sqrt{2k^3}}\echi (\bfk),~~~
  \dphi (\bfk)=\frac{H(t_k)}{\sqrt{2k^3}}e^{\gkc(t_k)/2}\ephi (\bfk),
\eeq
where the exponential factor in the latter equality is present because
$\phi$ has a non-canonical kinetic term in the action in the Einstein
frame.  Here $\echi (\bfk)$ and $\ephi (\bfk)$ are classical
random Gaussian quantities with the following averages.
\beq
  \langle\echi (\bfk)\rangle=  \langle\ephi (\bfk)\rangle=0,~~~
  \langle\epsilon_i (\bfk)\epsilon^*_j(\bkp)\rangle
  =\delta_{ij}\delta^{(3)}(\bfk-\bkp),~~~i,j=\chi,\phi.
\eeq

We thus find
\beqa
C_1&=&\lkk e^{\gkc}H\frac{\dphi}{\dot{\phi}}
 +(1-e^{\gkc})H\frac{\dchi}{\dot{\chi}}\rkk_{t_k}
=\frac{H^2(t_k)}{\sqrt{2k^3}}
 \lkk \frac{e^{\frac{3}{2}\gkc}}{\dot{\phi}}\ephi (\bfk)
 +\frac{1-e^{\gkc}}{\dot{\chi}}\echi (\bfk)\rkk_{t_k},  \label{C1}\\
C_3&=&\lkk
e^{\gkc}H\lmk\frac{\dphi}{\dot{\phi}}-\frac{\dchi}{\dot{\chi}}\rmk\rkk_{t_k}
=\frac{H^2(t_k)}{\sqrt{2k^3}}
\lkk \frac{e^{\frac{3}{2}\gkc}}{\dot{\phi}}\ephi (\bfk)
-\frac{\echi (\bfk)}{\dot{\chi}}\rkk_{t_k}.
\eeqa
Note the interesting fact that in the specific case of the "soft"
inflation \cite{BMY} ($\gamma \equiv 0$), the Brans-Dicke dilaton
does not contribute to the adiabatic mode ($C_1$) at all.

\subsection{Adiabatic modes}
{}From (\ref{gruniv}) curvature perturbation due to primordially
adiabatic fluctuation is given by
\beq
  \Phi_{\rm ad}=\lkk 1+\frac{2}{3(1+w)}\rkk^{-1}C_1,
  ~~~w\equiv\frac{\rho}{p},
\eeq
with $p$ and $\rho$ being total pressure and energy density,
respectively.
since $\Phi$ is related to the gauge-invariant density fluctuation
$\delta\rho^{(c)}/\rho$,
which reduces to $\delta\rho/\rho$ in the comoving gauge, as
\beq
  \frac{\delta \rho^{(c)}}{\rho} = -\frac{2}{3}\lmk\frac{k}{Ha}\rmk^2\Phi,
\eeq
we find
\beq
  \frac{\delta \rho^{(c)}}{\rho}=
 f\lkk e^{\gkc}H\frac{\dphi}{\dot{\phi}}
 +(1-e^{\gkc})H\frac{\dchi}{\dot{\chi}}\rkk_{t_k},  \label{adfluc}
\eeq
at the horizon crossing ($k=aH$), where $f$ is a constant equal
to 4/9 during radiation domination and to 2/5 during matter domination
(we omit the minus sign here because it may be absorbed
into the stochastic variables $\delta \phi$ and $\delta \chi$).

The above formula is quite different from that used by the previous
authors \cite{BM,Mc,MM,Der,Garcia}, who did not account for
the isocurvature mode and its mixing with the adiabatic mode during
inflation,
and concluded that a density fluctuation at the second horizon
crossing is approximately equal to $\delta\rho/(\rho+p)$
evaluated at the first horizon crossing during inflation.  That is,
up to a factor of order of unity,
\beq
  \frac{\delta\rho}{\rho}\simeq \left.
  \frac{H(|\dot{\chi}|\dchi + e^{-\gkc}|\dot{\phi}|\dphi)}
  {\dot{\chi}^2+e^{-\gkc}\dot{\phi}^2}\right|_{t_k}
{\rm (incorrect),}
  \label{maeda}
\eeq
which gives
\beqa
\frac{\delta\rho}{\rho}\simeq \lnk \begin{array}{ll}
  \frac{H\dchi}{|\dot{\chi}|} &
  \mbox{~~~~~for $|\dot{\chi}|>|\dot{\phi}|e^{-\gkc/2}$}\\
  \frac{H\dphi}{|\dot{\phi}|} &
  \mbox{~~~~~for $|\dot{\chi}|<|\dot{\phi}|e^{-\gkc/2}$}\\
  \end{array} \right.
   \mbox{~~~~~(incorrect).}
\eeqa
Note also that we would have reached the above formula (\ref{maeda})
or its analogue in the Jordan frame \cite{Garcia}, had we put $C_3=0$
neglecting the isocurvature mode in (\ref{grfluc}).  Clearly this is
not justifiable because it results in imposing a redundant constraint
between $\dchi $ and $\dphi$ in eqs.\ (\ref{chifluc}) and (\ref{phifluc}).
The difference between the expressions (\ref{adfluc}) and (\ref{maeda})
is especially large near the points where $V'(\phi)=0$. Then Eq.
(\ref{adfluc}) predicts much larger adiabatic perturbations than
Eq. (\ref{maeda}).

We are interested in the amplitude of density fluctuations about
$40\sim60$ e-folds before the end of inflation, or at $z\simeq 40\sim
60$, corresponding to
large-scale structures within the horizon scale today.  From
(\ref{chisol}) the corresponding value of $\chi$ ranges
$\kappa\chi=-40\beta \sim -60\beta$, for which we have $e^{\gkc}\simeq
1.2$ with $\omega \simeq 500$.  Thus the above incorrect formula and
ours give practically opposite results with respect to the two cases
$|\dot{\chi}|\gsim |\dot{\phi}|$ and
$|\dot{\chi}|\lsim |\dot{\phi}|$.  In some models of inflation, unless
$\beta$ is too small, we find $|\dot{\phi}|<|\dot{\chi}|$, so that the
adiabatic fluctuation is given by
\beq
\frac{\delta\rho}{\rho} \simeq e^{\gkc}\frac{H\dphi}{|\dot{\phi}|},
\label{adphidom}
\eeq
which gives practically the same result as in the Einstein gravity.
On the other hand, in the limit $\omega \longrightarrow \infty$ or
$\beta,~\gamma \longrightarrow 0$, we find $e^{\gkc}=1$ so that the
second term in (\ref{adfluc}) does not contribute to adiabatic
fluctuation and we again find
\beq
\frac{\delta\rho}{\rho} \simeq \frac{H\dphi}{|\dot{\phi}|},
\eeq
in agreement with the result in the Einstein gravity.

\subsection{Isocurvature modes}

Isocurvature fluctuations can be significant if some ingredient of the
model is essentially decoupled from the usual matter and constitutes a
part of dark matter.  In our model, the Brans-Dicke dilaton is a
candidate of such an ingredient.

During inflation, its fractional comoving energy perturbation
satisfies the inequality,
\beq
\frac{\delta\rho_\chi^{(c)}}{\rho_\chi} <
\frac{2\delta\rho_\chi^{(c)}}{\rho_\chi+p_\chi}
=2\frac{\partial~}{\partial t}\lmk\frac{\dchi}{\dot{\chi}}\rmk -2\Phi
=\frac{C_3}{\kappa^2}\lmk\frac{V'(\phi)}{V(\phi)}\rmk^2  < C_3.
\eeq
Since $C_3$ is of the same order of $C_1$, which gives the amplitude
of adiabatic perturbation at the second horizon crossing, fractional
density fluctuation in $\chi$ is smaller than the adiabatic one.
Furthermore, its contribution to the total energy density fluctuation,
$\delta\rho_\chi^{(c)}/\rho$ turns out to be much smaller
than the adiabatic counterpart, because we find $\rho_\chi
=\order(\omega^{-1})\rho \ll \rho$
in the post inflationary universe as will be seen below.
We, therefore, conclude that isocurvature fluctuations are negligible
in this model.

\subsection{Density fluctuations in the physical (Jordan) frame}

We can relate the amplitude of density fluctuations in the Einstein
frame with that in the Jordan frame using the conformal transformation
\cite{Hwang}.
First we note that gauge-invariant variables in both frames are related
with each other as
\beq
  \Phihat=\Phi-\frac{\dOmega}{\Omega},~~~~~
  \Psihat=\Psi+\frac{\dOmega}{\Omega},
\eeq
where $\dOmega = \frac{\gamma\kappa}{2}\dchi\Omega$ is gauge-invariant
perturbation of $\Omega$.  Using the equalities
\beq
\ahat=\Omega^{-1}a,~~~d\that=\omega^{-1}dt,~~~
\Hhat=\frac{\ahat_{\that}}{\ahat}=\Omega\lmk
H-\frac{\dot{\Omega}}{\Omega}\rmk,
\eeq
one can write (\ref{gruniv}) and (\ref{scalaruniv}) in terms of
physical variables as
\beqa
 \Phi&=&C_1\lkk 1-\lmk\Hhat + \frac{\Omega_{\that}}{\Omega}\rmk
\frac{1}{\ahat\Omega^2}\int^{\that}\ahat\Omega^2d\that ' \rkk  \\
 \frac{\dchi}{\dot{\chi}}&=&\Omega\frac{\dchi}{\chi_{\that}}
 =\frac{C_1}{\ahat\Omega}\int^{\that}\ahat\Omega^2d\that ' .
\eeqa
We, therefore, find
\beqa
\Phihat&=&C_1\lkk 1-\lmk\Hhat + 2\frac{\Omega_{\that}}{\Omega}\rmk
\frac{1}{\ahat\Omega^2}\int^{\that}\ahat\Omega^2d\that ' \rkk , \\
\Psihat&=&C_1\lkk 1-
\frac{\Hhat}{\ahat\Omega^2}\int^{\that}\ahat\Omega^2d\that ' \rkk .
\eeqa
The above two equations are the desired formula to relate density
fluctuations in the Einstein frame and those in the Jordan frame
through $C_1$.
In practice, however, since the Brans-Dicke field varies extremely slowly
in the post-inflationary universe, one can regard $\Omega$ as a constant
in these equations.  For example, in the matter dominated stage, we
find \cite{BD}
\beqa
\Phibd(\that)&=&\Phi_{\rm BD0}\lmk\frac{\that}{\that_0}\rmk
^{\frac{2}{3\omega+4}},  \label{bdmd} \\
\ahat (\that )&=&\ahat_0\lmk\frac{\that}{\that_0}\rmk
^{\frac{2+2\omega^{-1}}{3+4\omega^{-1}}},
\eeqa
Hence $\Omega_{\that}/\Omega$ is smaller than $\Hhat$ by a factor
less than $1/500$.  Thus we may simply conclude that the adiabatic
fluctuations are described by the same formula in both frames.

Using (\ref{bdmd}), we can estimate energy density of $\chi$ in the
Einstein frame in this stage as
\beq
\rho_\chi = \frac{1}{2}\dot{\chi}^2=\frac{2\omega+3}{(3\omega+5)^2}
\frac{1}{\kappa^2t^2}\cong \frac{1}{6\omega}\rho .
\eeq
Thus we may conclude the isocurvature perturbation due to $\chi$
is not important as stated in the previous subsection.

\section{Application to specific inflationary models}
Having developed general considerations, we now apply the above results
to two specific inflation models, namely chaotic inflation
\cite{chaoinf} and new inflation \cite{newinf},
and examine constraints on their model parameters as well
as the spectral shape.

\subsection{Chaotic inflation}
Here we consider two different potentials
$V(\phi)=\frac{1}{2}m^2\phi^2$ and $V(\phi)=\frac{\lambda_4}{4}\phi^4$
which are denoted collectively as $V(\phi)=\frac{\lambda_n}{n}\phi^n$.
In this model, inflation occurs at large $\phi$ and it is terminated
when $|\dot{\phi}/\phi|$ becomes as large as $H$ at $\phi=\phi_f=
\sqrt{n}/\kappa$.  Since we are interested in the last 60 e-foldings
in the inflationary stage, we may approximate (\ref{phisol}) as
\beq
  \kappa^2\int_{\phi_f}^{\phi}\frac{V(\varphi)}{V'(\varphi)}d\varphi
 =\frac{\kappa^2}{2n}\lmk\phi^2-\frac{n}{\kappa^2}\rmk \simeq z.
\eeq
The error in the last expression is only about $12\%$ with $z=60$,
$\beta=2\gamma=0.09$.
The e-folding number, $\zk$, when the comoving wave-number $k$ leaves
the Hubble radius during inflation satisfies
\beq
  \frac{k}{k_f}=e^{\lmk 1-\frac{\beta^2}{2}\rmk\zk}
  (2\zk)^{\frac{n}{4}},~~~~~\mbox{for}~~ 1 \ll \zk \lsim 60.
\eeq
We can express the amplitude of curvature perturbation on scale
$l=\frac{2\pi}{k}$ as
\beqa
\Phihat(l)&=&\lkk 1+\frac{2}{3(1+w)}\rkk^{-1}
\frac{\sqrt{2k^3\langle |C_1|^2 \rangle}}{2\pi} \nonumber \\
&=&\lkk 1+\frac{2}{3(1+w)}\rkk^{-1}
\frac{\kappa^2}{2\pi}\sqrt{\frac{\lambda_n}{3n}
\lmk \frac{2n\zk}{\kappa^2}\rmk^{\frac{n}{2}}}
\lkk e^{\frac{1}{2}\beta(\gamma-\beta)\zk}
\sqrt{\frac{2\zk}{n}}
+\frac{e^{-\frac{\beta^2}{2}\zk}}{\beta}(e^{\beta\gamma\zk}-1) \rkk,
\eeqa
again for $1 \ll \zk \lsim 60$.

Since the large-angular-scale anisotropy of background radiation due to
the Sachs-Wolfe effect is given by $\delta T/T = \Phihat/3$,
we can normalize the value of $\lambda_n$ by the COBE observation \cite{COBE}.
For $\beta=2\gamma=0.09$, we find
\beqa
\frac{\delta T}{T} &=& \frac{1}{3}\Phihat (\zk \simeq 60)
\simeq \lnk \begin{array}{ll}
  9\frac{m}{\mpl} &
  \mbox{~~~~~for $n=2$}\\
  32\sqrt{\lambda_4}&
  \mbox{~~~~~for $n=4$}\\
  \end{array} \right. \\
  &=& 1.1\times 10^{-5}
\eeqa
Note that our normalization of the background solution $a_f=1,~
\chi_f=0$ (see Eqs. (\ref{chisol},\ref{phisol})) means that $m$ and
$\lambda_4$ are values of physical parameters ("coupling constants")
at the end of inflation. However, since $\chi$ generally grows only
as logarithm of $t$ after inflation and is even constant during the whole
radiation-dominated stage in the Brans-Dicke case $\beta=2\gamma$,
their present values are not significantly different from those at the
end of inflation.  So, it is natural to use just the latter values
when comparing non-Einstein gravity models having variable
coupling constants with models based on the Einstein gravity. We find
\beqa
  m&=& 1\times 10^{13} \mbox{GeV},~~~~~n=2, \\
  \lambda_4 &=& 1\times 10^{-13},~~~~~n=4
\eeqa
which is no different from the values obtained assuming the Einstein
gravity \cite{Salopek}.
Since the behavior of the system approaches to that in the
Einstein gravity as we increase $\omega$, we can conclude that in
Brans-Dicke theory, model
parameters of the inflaton's potential should take the same value as
in the Einstein gravity.


\subsection{New inflation}
Next we consider new inflation with a potential
\beq
  V(\phi)=V_0 -\frac{\lambda}{4}\phi^4,
\eeq
for which we find, from (\ref{phisol}),
\beq
  \kappa^2\int_{\phi_f}^{\phi}\frac{V(\varphi)}{V'(\varphi)}d\varphi
\cong \frac{\kappa^2V_0}{2\lambda}\lmk \frac{1}{\phi^2}
-\frac{1}{\phi_f^2}\rmk \cong \frac{\kappa^2}{2\lambda\phi^2}
\simeq z.
\eeq
We can again express the amplitude of curvature fluctuation as a
function of $\zk$, which is now related with $k$
as,
\beq
\frac{k}{k_f}=e^{\lmk 1-\frac{\beta^2}{2}\rmk\zk}H_f.
\eeq
\beq
\Phihat(l)=\lkk 1+\frac{2}{3(1+w)}\rkk^{-1}
\lkk e^{-\frac{1}{2}\beta(\beta-\gamma)\zk}
\sqrt{\frac{\lambda}{3}}(2\zk)^{\frac{3}{2}} +
\kappa H_f \frac{e^{-\frac{\beta^2}{2}\zk}}{\beta}
(e^{\beta\gamma\zk}-1)\rkk,
\eeq
with $H_f\equiv \sqrt{\frac{\kappa^2}{3}V_0}$.
Taking $\beta=2\gamma=0.09$ again, it predicts the amplitude of
$\delta T/T$ to be compared with COBE data as
\beq
\frac{\delta T}{T}(\zk \simeq 60)
\simeq 21\sqrt{\lambda} +0.4\frac{H_f}{\mpl}.  \label{newinfdT}
\eeq
Since $H_f$ should  also satisfy
\beq
\frac{H_f}{\mpl} \lsim 10^{-5}
\eeq
to suppress long-wave gravitational radiation of quantum origin \cite{GW},
we find
\beq
  \lambda \lsim 2\times 10^{-13}
\eeq
from (\ref{newinfdT}).
Again its amplitude is practically no different from the case of
the Einstein gravity.


\section{Conclusion}

In the present paper we have investigated density perturbations generated
during inflation in the Brans-Dicke theory, paying attention to the fact
that quantum fluctuations of both the inflaton and the Brans-Dicke
dilaton fields contribute
to them. This implies that not only adiabatic (curvature) but also
isocurvature perturbations exist and that the simple conservation
 formula of a perturbation outside the Hubble radius cannot be used.

Working in the Einstein frame, which is connected with the original
frame through a conformal transformation, we have obtained expressions
for growing adiabatic and isocurvature modes and have shown that
only the former is important after inflation.  Then we have corrected
the previously
used formula based on the use of incorrect matching of super-horizon
perturbations with quantum fluctuations at horizon (Hubble radius)
crossing.
Since both the inflaton and the dilaton turn out to contribute to
the perturbation with the same order of magnitude, the
amplitude itself is not significantly altered compared with the case
of inflation in the Einstein gravity.

However, our constraints on the model parameters turned out to be
more severe than those obtained by Berkin and Maeda \cite{BM}.
This is because
they have taken the initial value of $\chi$ rather arbitrary, which
might be appropriate to the case of original soft inflation model in
which inflaton's potential is multiplied by an exponential
potential by hand \cite{BMY}.
On the other hand,
we, concentrating on the Brans-Dicke model, had to normalize its
value to be zero after inflation to reproduce the correct gravitational
constant today.

In conclusion, values of inflaton's coupling parameters in the Brans-Dicke
gravity (if taken at the end of inflation) which are required to
produce the correct amplitude of present adiabatic perturbations
are practically no different from those in the Einstein gravity
in the case of both chaotic and new inflationary models.

\acknowledgements
\noindent
A.\ S.\ is grateful to Profs.\ Y.\ Nagaoka and J.\ Yokoyama for their
hospitality at the Yukawa Institute for Theoretical Physics, Kyoto
University where this project was started. A.\ S.\ was supported in part
by the Russian Foundation
for Basic Research, Project Code 93-02-3631, and by Russian Research
Project ``Cosmomicrophysics''.  J.\ Y.\ acknowledges support by
the Japanese Grant-in-Aid for
Scientific Research Fund of Ministry of Education, Science, and
Culture, No.\ 06740216.

\end{document}